\documentclass[conference]{IEEEtran}
\IEEEoverridecommandlockouts
\usepackage{cite}
\usepackage{amsmath,amssymb,amsfonts}
\usepackage{algorithmic}
\usepackage{graphicx}
\usepackage{textcomp}
\usepackage{xcolor}
\def\BibTeX{{\rm B\kern-.05em{\sc i\kern-.025em b}\kern-.08em
    T\kern-.1667em\lower.7ex\hbox{E}\kern-.125emX}}

\usepackage[font=small,labelfont=bf,tableposition=top]{caption}

\usepackage{blindtext}

\title{Facial Emotion Recognition in VR Games\\
}
\author{\IEEEauthorblockN{Fatemeh Dehghani}
\IEEEauthorblockA{\textit{Faculty of Business and Information Technology} \\
\textit{Ontario Tech University}\\
Oshawa, Canada \\
Fatemeh.Dehghani@ontariotechu.ca}
\and
\IEEEauthorblockN{Loutfouz Zaman}
\IEEEauthorblockA{\textit{Faculty of Business and Information Technology} \\
\textit{Ontario Tech University}\\
Oshawa, Canada \\
loutfouz.zaman@ontariotechu.ca}
}
\begin{document}
\IEEEoverridecommandlockouts
\IEEEpubid{\makebox[\columnwidth]{979-8-3503-2277-4/23/\$31.00~\copyright2023 IEEE \hfill} 
\hspace{\columnsep}\makebox[\columnwidth]{ }}
\maketitle
\IEEEpubidadjcol
\begin{abstract}
Emotion detection is a crucial component of Games User Research (GUR), as it allows game developers to gain insights into players' emotional experiences and tailor their games accordingly. However, detecting emotions in Virtual Reality (VR) games is challenging due to the Head-Mounted Display (HMD) that covers the top part of the player's face, namely, their eyes and eyebrows, which provide crucial information for recognizing the impression. To tackle this we used a Convolutional Neural Network (CNN) to train a model to predict emotions in full-face images where the eyes and eyebrows are covered. We used the FER2013 dataset, which we modified to cover eyes and eyebrows in images. The model in these images can accurately recognize seven different emotions which are anger, happiness, disgust, fear, impartiality, sadness and surprise.\\

We assessed the model's performance by testing it on two VR games and using it to detect players' emotions. We collected self-reported emotion data from the players after the gameplay sessions. We analyzed the data collected from our experiment to understand which emotions players experience during the gameplay. We found that our approach has the potential to enhance gameplay analysis by enabling the detection of players' emotions in VR games, which can help game developers create more engaging and immersive game experiences.
\end{abstract}
\begin{IEEEkeywords}
Players, Emotions, Virtual Reality, Games, Facial Expressions.
\end{IEEEkeywords}
\section{Introduction}
Virtual Reality (VR) refers to simulated environments in which physical properties of the real world can be emulated. VR is used in diverse domains, such as medicine, social sciences, psychology, and entertainment. One of the most used applications of VR is in video games, which is one of the world’s most popular and prevailing entertainment tools, offering game players an innovative gaming experience.\\

Game analytics involves utilizing analytics for game development and research purposes\cite{drachen2013game}. In game analytics data mining methods are used to extract insights from game-related data and for pattern detection, with a particular focus on player behavior \cite{wallner2019data}. The process of examining data goes through data preparation, modeling, and evaluation, which helps game developers and game researchers to improve games to the extent that game players catch the experience that designers planned for \cite{maccormick2020echo}. To achieve this, game developers have designed many evaluation methods to identify problems that a player might face while playing and then resolve these problems using player data. The level of player engagement is considered a primary key in evaluating the success of games. In order to achieve this goal, game developers look for toolkits and methods to understand why players give up on the specific level of computer games and never return to finish them \cite{schoenau2011player}. One of the ways to figure out the reason for this is analyzing players’ impressions. It has been observed that players show different reactions, and experience both negative and positive emotions (e.g., sadness, fear, excitement, bored) while facing challenges \cite{yannakakis2014emotion}. Understanding players' emotions is very crucial for game developers as it helps them to 
\begin{itemize}
  \item understand the level of player engagement,
  \item ensure positive emotional states,
  \item introduce the right level of dynamism to the game\cite{burns2017detecting}.
\end{itemize}
If the player's expression suggests stress, the pertinent game difficulty factors may be lowered until the player is at ease \cite{akbar2019enhancing}.

\section{Related work}
Burns and Tulip \cite{burns2017detecting} developed a method to measure gamer involvement, specifically "flow," using facial expressions. They recorded players' facial expressions while gaming with a webcam and used machine learning to identify features associated with flow, achieving 78\% accuracy on their dataset. Liu et al. \cite{Liu2009DynamicDA} conducted research on the use of real-time anxiety-based feedback to implement Dynamic Difficulty Adjustment (DDA) in games. They proposed tracking players' physiological reactions, particularly their skin conductance levels, to measure fear and adjust the game difficulty accordingly to maintain a state of flow. A study with 15 individuals playing a modified version of \emph{Tetris} found that DDA improved player engagement and positive affective experiences compared to playing without DDA. Khezri et al. \cite{khezri2015reliable} proposed a method to improve emotion recognition by combining multiple emotional modalities. They used recorded signals and multiple classification units to independently identify emotions, resulting in a significant performance boost for their system. Yang et al. \cite{yang2018affective} introduced a method to classify emotions according to skin conductance and electroencephalography signals. The classification was performed on images. The method establishes a connection between experiential information and the viewer's expected emotion experience, resulting in a positive experience. Jang et al. \cite{jang2019reliability} checked the accuracy of using changes in physiological signals in response to emotions. Six emotions were measured. Jang et al. found skin conductance, blood volume signals and heart rate more dependable compared to baseline evaluations. Ouellet \cite{Ouellet2014RealtimeER} proposed a new technique for recognizing emotions in real-time gaming using a Convolutional Neural Network (CNN) to analyze facial expressions of players. Players' faces were detected with the Viola-Jones detector in OpenCV from a continuous video stream captured from the webcam. The proposed method was compared to other emotion recognition methods and showed better accuracy and speed, with a 94.4\% accuracy rate. Hickson et al.\cite{8658392} described a unique method for identifying and categorizing facial expressions related to five different emotions by utilizing eye movement and gaze attributes. Long Short-Term Memory (LSTM) networks are combined with CNNs to achieve 74\% accuracy in recognizing face expressions and 70\% accuracy in detecting facial action units. Suzuki et al.\cite{suzuki2017recognition} developed a technique to use integrated photo reflecting sensors in an HMD to identify and map face emotions to avatars. Their method recognized seven fundamental facial emotions with an accuracy of 82\% and mapped these expressions to avatars with an accuracy of 86\%.\\

We avoid the complexity, expenses, and possible constraints associated with eye-tracking and sensor gear by concentrating entirely on facial emotions collected through full-face photos. Our technique provides generalizability across multiple VR systems and gaming conditions, allowing us to accommodate instances where eye-tracking may not be provided or maintained continuously by players.\\

CNNs demonstrated success in image processing since their inception in the late 1990s \cite{lecun1995convolutional}. The architecture of a CNN makes it particularly effective for handling static images \cite{sri2022facial}. In recent years, larger datasets and increased processing power have made CNNs a more practical approach for feature extraction and image categorization \cite{10.1145/3065386}.

\section{Proposed Method}
In this work, we aim to assess the emotions of players engaged in VR gaming while wearing HMDs. These displays tend to occlude the upper facial regions, such as the eyes and eyebrows, known for providing crucial cues in emotion recognition. We first created our required dataset: face images with covered eyes and eyebrows. Next, we trained the existing facial emotion recognition model on the original dataset, FER, and then on our dataset, to see if the model can estimate human emotions based on the mouth and chin. We then performed an evaluation. Next, in order to see the real application of this work, we asked five participants to play two VR games, and recorded their faces while wearing the VR headset. We also recorded their gameplay synchronously. After each game, participants filled out a form where they were asked which emotions they experienced while playing the games. We then converted the videos to images and fed them to the model. We compared the feelings declared by participants in each game and the emotions predicted by the model to see how accurately the model classifies emotions. \\
\subsection{VGGNet}
VGGNet is an architecture of convolutional neural networks used in pattern recognition and image processing\cite{simonyan2014very}. Four convolutional blocks are responsible for extracting the features of high level, and the task of categorizing emotions is performed by the fully connected layers \cite{khaireddin2021facial}.

\subsection{Dataset}

\begin{figure}[h]
  \centering
  \includegraphics[width=0.6\linewidth]{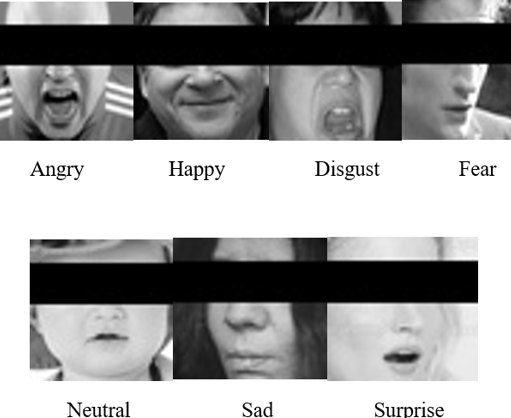}
  \caption{Samples of all the expressions with eyes and eyebrows blacked out for training the model.}
  \label{fig:new_data_set}
\end{figure}

The FER2013 is a publicly accessible dataset of facial emotion recognition (FER), which contains a vast collection of 35,887 face crops. The dataset has train, test and validation sets. For each sample, simple expression labels are offered. Each image has a dimension of 48 × 48 pixels and is grayscale. FER2013 provides seven emotional expressions: happy, fearful, disgusted, sad, angry, neutral, and surprised. Moreover, we recorded ten 20-minute long videos, two per participant (one for each game), while participants were playing, then converted them to images and made annotations to make them ready to feed the model. For this purpose, we got the pixels of images and saved them as a CSV file, and then fed them to the model, which was trained on the covered eyes dataset (Fig. \ref{fig:new_data_set}), to see which emotions the model can accurately detect.

\section{Training and testing of the model}
To create our desired dataset containing a covered-eye human face, we used the OpenCV library to get the coordinates of eyes and eyebrows in the images. Then, we rendered a black rectangle that covered these areas. We applied this function on 35{,}000 images. Eyes and eyebrows were well covered in the images. Then, since the model's checkpoint was unavailable, we first trained the model on the original dataset over 300 epochs. Next, we trained the model on our desired dataset.
The model used in this work achieved a 73.28\% accuracy of a single-network on FER2013. Additional training data was not utilized. When tested on the covered eyes dataset, the model's accuracy was 68.59\%. \\

\section{Real World Validation}
While our model showed good results in theory, we also wanted to put it into practice and see if it is actually usable as a tool for emotion detection in Games User Research (GUR). For this validation, we designed a small user study where we followed a traditional Human-Computer Interaction (HCI) procedure described below. For this study, we picked two PSVR games: \emph{Astro Bot Rescue Mission} and \emph{Paranormal Activity}. The reason they were chosen is because we hypothesized that together these two games would be able to capture most  emotions our model is capable of detecting. Namely, we hypothesized that playing \emph{Astro Bot Rescue Mission} would result in detecting positive emotions, while \emph{Paranormal Activity} would result in negative ones. 

\subsection{Participants}
We recruited five participants (three males and two females) to play two games: \emph{Astro Bot Rescue Mission} and \emph{Paranormal Activity}. The participants were between the ages of 21 and 27 (\emph{Mdn} = 24) and were recruited from the graduate student population in computer science at our university. We gathered participants' demographic information through a Google form. Participants expressed whether they played VR games before. While three had played VR games before, two participants did not. Also, they were asked to rate their expertise as video game players. One considered themselves a casual player, three consider themselves core players, and one was a hardcore player. Moreover, they were asked about their familiarity with the games. Two were new to \emph{Astro Bot Rescue Mission}, and three were new to \emph{Paranormal Activity}.\\

\subsection{Apparatus}
\begin{figure}[h]
  \centering
  \includegraphics[width=0.8\linewidth]{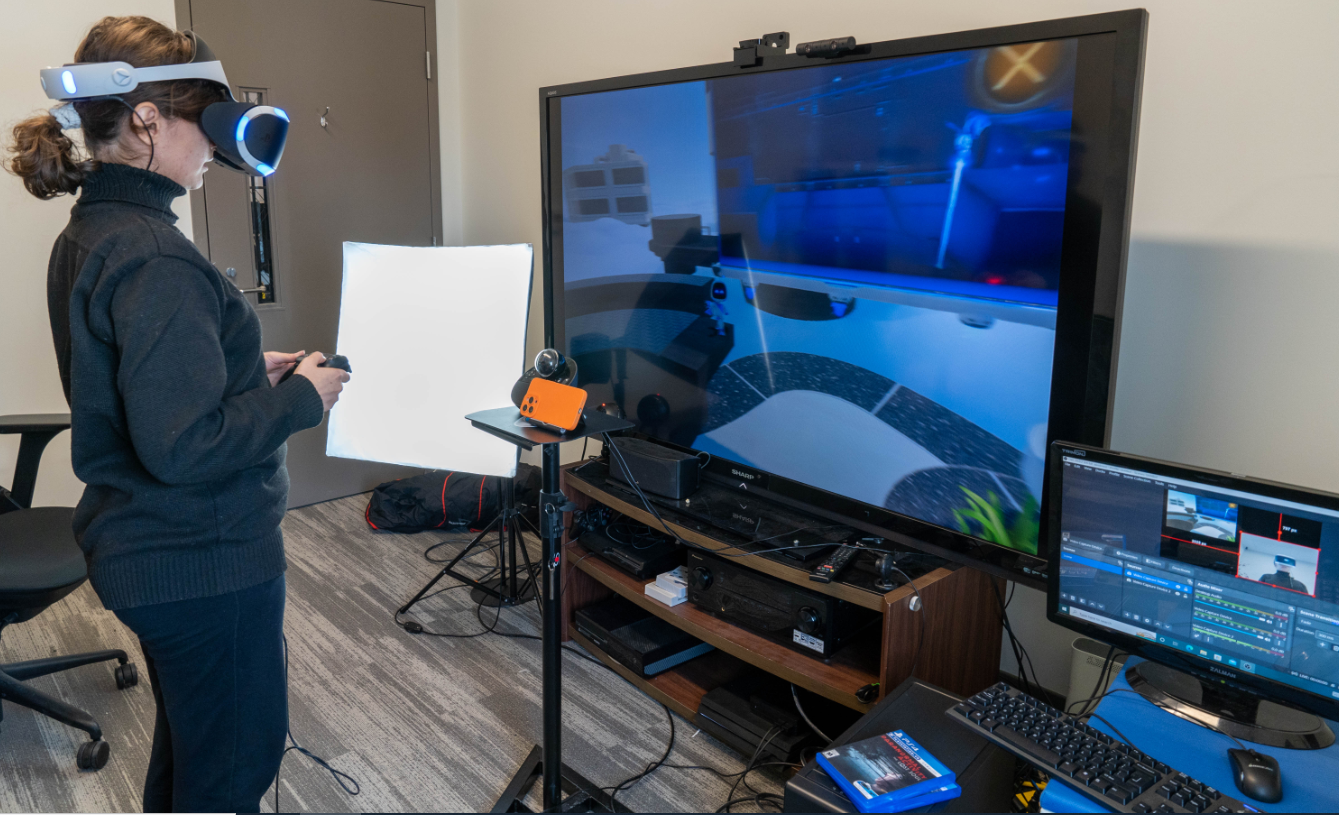}
  \caption{Environment and equipment used for collecting the data.}
  \label{fig:study_setup}
\end{figure}

 We used two camera devices to record players' faces, one of which was Apple \emph{iPhone 14 Pro Max} and the other Logitech \emph{BCC950 ConferenceCam}. The gameplay was captured using \emph{4K Ultra HD HDMI Capture} card. The \emph{ConferenceCam} footage was synced with the output of the capture card and recorded using Open Broadcaster Software (OBS) during the gameplay for later analysis. Both cameras were placed  at about waist height to the participants and facing upwards to capture the facial expression under the VR helmet. A PC was used to record the output from the \emph{ConferenceCam} and the capture card. The \emph{iPhone} footage was used for feeding into the model because it produced video quality that was more acceptable for this purpose, unlike that of the Logitech camera. Both games were hosted on Sony \emph{PlayStation 4} with \emph{PSVR}. The players used \emph{PlayStation 4} controllers for playing the games. Fig \ref{fig:study_setup}
shows the environment and equipment we used for collecting
the data.

\subsection{Procedure}
After filling out the demographic questionnaire, we instructed the participants to play one of the two games first. We counterbalanced the order in which the games were presented to the participants. The participants played the game from the beginning. In \emph{Astro Bot Rescue Mission} participants started from World 1-1 and played until the 20 minute time limit expired. In \emph{Paranormal Activity} participants played the game from the very beginning and had to perform all the steps to enter the house and pass through the first door. When stuck they received verbal instructions from the 1st author of this work. After the participants played the game for 20 minutes, they were requested to complete an in-game survey, where on a 5-point Likert scale they rated how much anger, happiness, disgust, fear, neutrality, happiness, sadness, and surprise they experienced, with 1 being not experiencing the corresponding emotion at all, and 5 being experiencing it very much. The procedure was then repeated for the other game. The whole study took about 1 hour.

\section{Results}
In total, we recorded 5 videos for each game (10 in total). We converted the videos to images. We then obtained their pixels and saved them in two separate CSV files to detect experienced emotions in each game. We put an empty column in the files for the model to predict corresponding labels, indicating detected emotions in images. After that, we compared the detected feelings by the model and the declared emotions by each player. Furthermore, we reviewed the recorded videos alongside the forms filled out by participants after each game for validation.\\

In terms of emotion detection, participants reported experiencing three emotions: happy, neutral, and surprise, while playing \emph{Astro Bot Rescue Mission}. The model successfully detected happy and neutral emotions during this game. When playing  \emph{Paranormal Activity}, participants reported experiencing four emotions: anger, fear, surprise, and neutral, and the model was able to detect fear, neutral, and surprise emotions accurately. In \emph{Astro Bot Rescue Mission}, participants felt happy mostly when all rescued bots appeared and danced before the end of each level or sometimes when the bot waved its hands. In \emph{Paranormal Activity}, most participants felt fear 2--3 times: when they walked by an opened window and it suddenly closed, when they went upstairs, and the girl (NPC character) suddenly came downstairs, when they saw the lady with an axe in her hand on the second floor in the house, which was the scariest scene for all of them. Some of them were surprised when they saw the TV screen showing the girl (NPC character) scared of seeing something or when objects moved. The neutral feeling was the most prevailing emotion in both games.\\

\begin{figure}[h]
  \centering
  \includegraphics[width=\linewidth]{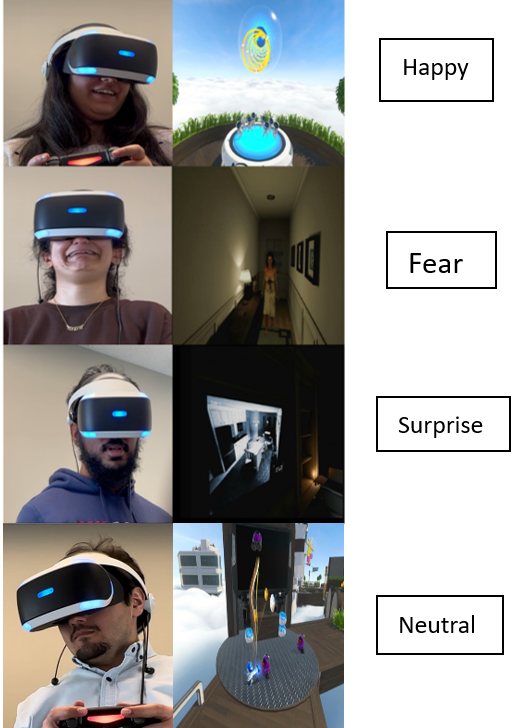}
  \caption{Some of the pictures from the video footage show participants' faces, the game scene, and the detected emotions.}
  \label{fig:samples}
\end{figure}

Overall, our experimental findings indicate that the CNN can accurately categorize emotions using solely the facial features extracted from areas of the face other than the eyes and eyebrows, even when the facial information in those regions is absent. This suggests that the model can effectively detect emotions in different game scenarios. Some of the pictures from the video footage show participants' faces, the game scene, and the detected emotions. See Fig. \ref{fig:samples}. 

\section{Limitations and Future Work}
 The obvious limitation of the current work is that we did not capture players' expressions when they moved their heads or turned around while playing. The model could only detect emotions when players' faces were facing the camera straight. This limitation can be addressed by attaching, e.g., a selfie stick to the VR helmet, so that the camera is always facing the player's face as they turn their head. Alternatively, the model can be fine-tuned in the future to detect emotions from multiple views. One aspect for further consideration is ensuring precise synchronization between the recorded emotions and the specific times they occurred, in order to enhance confidence in the accuracy and validation of our model. Furthermore, an emotion-based difficulty adjustment solution for VR games can be developed based on this work, which potentially could improve player experience.

\bibliographystyle{IEEEtran}
\bibliography{name}

\begin{thebibliography}{10}
\providecommand{\url}[1]{#1}
\csname url@samestyle\endcsname
\providecommand{\newblock}{\relax}
\providecommand{\bibinfo}[2]{#2}
\providecommand{\BIBentrySTDinterwordspacing}{\spaceskip=0pt\relax}
\providecommand{\BIBentryALTinterwordstretchfactor}{4}
\providecommand{\BIBentryALTinterwordspacing}{\spaceskip=\fontdimen2\font plus
\BIBentryALTinterwordstretchfactor\fontdimen3\font minus
  \fontdimen4\font\relax}
\providecommand{\BIBforeignlanguage}[2]{{%
\expandafter\ifx\csname l@#1\endcsname\relax
\typeout{** WARNING: IEEEtran.bst: No hyphenation pattern has been}%
\typeout{** loaded for the language `#1'. Using the pattern for}%
\typeout{** the default language instead.}%
\else
\language=\csname l@#1\endcsname
\fi
#2}}
\providecommand{\BIBdecl}{\relax}
\BIBdecl

\bibitem{drachen2013game}
A.~Drachen, M.~Seif El-Nasr, and A.~Canossa, ``Game analytics--the basics,''
  \emph{Game analytics: Maximizing the value of player data}, pp. 13--40, 2013.

\bibitem{wallner2019data}
G.~Wallner, \emph{Data Analytics Applications in Gaming and
  Entertainment}.\hskip 1em plus 0.5em minus 0.4em\relax CRC Press, 2019.

\bibitem{maccormick2020echo}
D.~MacCormick and L.~Zaman, ``Echo: Analyzing gameplay sessions by
  reconstructing them from recorded data,'' in \emph{Proceedings of the Annual
  Symposium on Computer-Human Interaction in Play}, 2020, pp. 281--293.

\bibitem{schoenau2011player}
H.~Schoenau-Fog \emph{et~al.}, ``The player engagement process-an exploration
  of continuation desire in digital games.'' in \emph{Digra conference}, 2011.

\bibitem{yannakakis2014emotion}
G.~N. Yannakakis and A.~Paiva, ``Emotion in games,'' \emph{Handbook on
  affective computing}, vol. 2014, pp. 459--471, 2014.

\bibitem{burns2017detecting}
A.~Burns and J.~Tulip, ``Detecting flow in games using facial expressions,'' in
  \emph{2017 IEEE Conference on Computational Intelligence and Games
  (CIG)}.\hskip 1em plus 0.5em minus 0.4em\relax IEEE, 2017, pp. 45--52.

\bibitem{akbar2019enhancing}
M.~T. Akbar, M.~N. Ilmi, I.~V. Rumayar, J.~Moniaga, T.-K. Chen, and
  A.~Chowanda, ``Enhancing game experience with facial expression recognition
  as dynamic balancing,'' \emph{Procedia Computer Science}, vol. 157, pp.
  388--395, 2019.

\bibitem{Liu2009DynamicDA}
C.~Liu, P.~Agrawal, N.~Sarkar, and S.~Chen, ``Dynamic difficulty adjustment in
  computer games through real-time anxiety-based affective feedback,''
  \emph{International Journal of Human–Computer Interaction}, vol.~25, pp.
  506 -- 529, 2009.

\bibitem{khezri2015reliable}
M.~Khezri, M.~Firoozabadi, and A.~R. Sharafat, ``Reliable emotion recognition
  system based on dynamic adaptive fusion of forehead biopotentials and
  physiological signals,'' \emph{Computer methods and programs in biomedicine},
  vol. 122, no.~2, pp. 149--164, 2015.

\bibitem{yang2018affective}
M.~Yang, L.~Lin, and S.~Milekic, ``Affective image classification based on user
  eye movement and eeg experience information,'' \emph{Interacting with
  Computers}, vol.~30, no.~5, pp. 417--432, 2018.

\bibitem{jang2019reliability}
E.-H. Jang, S.~Byun, M.-S. Park, and J.-H. Sohn, ``Reliability of physiological
  responses induced by basic emotions: A pilot study,'' \emph{Journal of
  Physiological Anthropology}, vol.~38, no.~1, pp. 1--12, 2019.

\bibitem{Ouellet2014RealtimeER}
S.~Ouellet, ``Real-time emotion recognition for gaming using deep convolutional
  network features,'' \emph{ArXiv}, vol. abs/1408.3750, 2014.

\bibitem{8658392}
S.~Hickson, N.~Dufour, A.~Sud, V.~Kwatra, and I.~Essa, ``Eyemotion: Classifying
  facial expressions in vr using eye-tracking cameras,'' in \emph{2019 IEEE
  Winter Conference on Applications of Computer Vision (WACV)}, 2019, pp.
  1626--1635.

\bibitem{suzuki2017recognition}
K.~Suzuki, F.~Nakamura, J.~Otsuka, K.~Masai, Y.~Itoh, Y.~Sugiura, and
  M.~Sugimoto, ``Recognition and mapping of facial expressions to avatar by
  embedded photo reflective sensors in head mounted display,'' in \emph{2017
  IEEE Virtual Reality (VR)}.\hskip 1em plus 0.5em minus 0.4em\relax IEEE,
  2017, pp. 177--185.

\bibitem{lecun1995convolutional}
Y.~LeCun, Y.~Bengio \emph{et~al.}, ``Convolutional networks for images, speech,
  and time series,'' \emph{The handbook of brain theory and neural networks},
  vol. 3361, no.~10, p. 1995, 1995.

\bibitem{sri2022facial}
K.~S. Sri, N.~N. Kumar, and V.~D. Satish, ``Facial emotion recognition using
  dcnn algorithm,'' in \emph{2022 7th International Conference on Communication
  and Electronics Systems (ICCES)}.\hskip 1em plus 0.5em minus 0.4em\relax
  IEEE, 2022, pp. 1336--1341.

\bibitem{10.1145/3065386}
\BIBentryALTinterwordspacing
A.~Krizhevsky, I.~Sutskever, and G.~E. Hinton, ``Imagenet classification with
  deep convolutional neural networks,'' \emph{Commun. ACM}, vol.~60, no.~6, p.
  84–90, may 2017. [Online]. Available: \url{https://doi.org/10.1145/3065386}
\BIBentrySTDinterwordspacing

\bibitem{simonyan2014very}
K.~Simonyan and A.~Zisserman, ``Very deep convolutional networks for
  large-scale image recognition,'' \emph{arXiv preprint arXiv:1409.1556}, 2014.

\bibitem{khaireddin2021facial}
Y.~Khaireddin and Z.~Chen, ``Facial emotion recognition: State of the art
  performance on fer2013,'' \emph{arXiv preprint arXiv:2105.03588}, 2021.

\end{thebibliography}
\end{document}